\documentclass[letterpaper, 10 pt, conference]{ieeeconf}  
\IEEEoverridecommandlockouts 

\usepackage{cite}

\usepackage{graphicx}
\usepackage{xcolor}
\usepackage{epstopdf}
\usepackage{mathtools}
\usepackage{fancyhdr}
\usepackage{subcaption}
\usepackage{amsfonts}
\usepackage{amssymb}
\usepackage[T1]{fontenc}
\usepackage{color}	
\usepackage{siunitx}
\usepackage{multirow}
\usepackage{placeins}
\usepackage{booktabs}
\usepackage{rotating}
\usepackage{array}
\usepackage{dsfont}
\usepackage{bbm}
\usepackage{bbold}
\usepackage[hyphens]{url}

\usepackage{tikz}

\usepackage{epsfig}
\usepackage{epstopdf}
\usepackage{algorithm} % for algorithms 
\usepackage{algorithmic} % for algorithms

\usepackage{xcolor}

\newcommand{\mc}[1]{\mathcal{#1}}
\newcommand{\R}{\mathbb{R}}
\newcommand{\N}{\mathbb{N}}
\newcommand{\bs}[1]{\boldsymbol{#1}}
\newcommand{\mf}[1]{\mathbf{#1}}

\newcommand{\tup}[1]{\textup{#1}}

\usepackage{mathtools}
\usepackage[acronym]{glossaries}
\newacronym{DR}{DR}{Demand Response}
\newacronym{DSO}{DSO}{Distribution System Operator}
\newacronym{TSO}{TSO}{Transmission System Operator}
\newacronym{BLO}{BLO}{BiLevel Optimization Problem}
\newacronym{GNE}{GNE}{Generalized Nash Equilibrium}
\newacronym{GNEP}{GNEP}{Generalized Nash Equilibrium Problem}
\newacronym{vGNE}{\textit{v}-GNE}{\textit{variational} Generalized Nash Equilibrium}
\newacronym{lSE}{$\ell$-SE}{local Stackelberg equilibrium}
\newacronym{MPEC}{MPEC}{Mathematical Program with Equilibrium Constraints}
\newacronym{MIP}{MIP}{Mixed-Integer Program}
\newacronym{KKT}{KKT}{Karush–Kuhn–Tucker}
\newacronym{IoT}{IoT}{Internet of Things}
\newacronym{DeePC}{DeePC}{Data-enabled Predictive Control }
\newacronym{MPC}{MPC}{Model Predictive Control }
\newacronym{MFD}{MFD}{Macroscopic Fundamental Diagram }
\newacronym{SUMO}{SUMO}{Simulation of Urban MObility}
\newacronym{CAV}{CAV}{Connected Autonomous Vehicle}

\newtheorem{lemma}{Lemma}

\newtheorem{remark}{Remark}

\newcommand{\Image}{\operatorname{im}} % image
\newcommand{\col}{\operatorname{col}}
\newcommand{\rank}{\operatorname{rank}} % rank
\newcommand {\T}{\mathbb{T}} 	% Time
\newcommand {\W}{\mathbb{W}} 	% Signals
\newcommand {\B}{\mathcal{B}} 	% Behavior
\renewcommand {\L}{\mathcal{L}} 	% Linear behaviors
\newcommand {\bma}{\left[}
\newcommand {\ema}{\right]}
\newcommand{\Tini}{{T_{\textup{ini}}}}
\newcommand{\Tf}{{T_{\textup{f}}}}
\newcommand{\uini}{{u_{\textup{ini}}}}
\newcommand{\yini}{{y_{\textup{ini}}}}
\newcommand{\wini}{{w_{\textup{ini}}}}

\newcommand{\Up}{{U_{\mathrm{p}}}}
\newcommand{\Uf}{{U_{\mathrm{f}}}}
\newcommand{\Yp}{{Y_{\mathrm{p}}}}
\newcommand{\Yf}{{Y_{\mathrm{f}}}}

\newcommand {\nn}{\nonumber}
\newcommand{\beq}{\begin{equation}}
\newcommand{\eeq}{\end{equation}}
\newcommand {\bseq}{\begin{subequations}}
\newcommand {\eseq}{\end{subequations}}
\title{\LARGE \bf
Urban traffic congestion control: a DeePC change}

\author{Alessio Rimoldi$^{\star}$, Carlo Cenedese$^{\star}$,
Alberto Padoan$^{\star}$, 
Florian D\"orfler$^{\star}$, 
John Lygeros$^{\star}$%
\thanks{$^{*}$Authors are with Automatic Control Laboratory, Department of Electrical Engineering and Information Technology,
        ETH Z\"urich, Physikstrasse 3 8092 Z\"urich, Switzerland
        {\tt\small\{ccenedese,apadoan,fdorfler,jlygeros\}@ethz.ch}.
}
\thanks{Research is supported by NCCR Automation and funded by the Swiss National Science
Foundation (grant number 180545).}%
}

\begin{document}

\maketitle
\thispagestyle{empty}
\pagestyle{empty}

% 
% Abstract
\begin{abstract} 
Urban traffic congestion remains a pressing challenge in our rapidly expanding cities, despite the abundance of available data and the efforts of policymakers. By leveraging behavioral system theory and data-driven control, this paper exploits the \acrshort{DeePC} algorithm in the context of urban traffic control performed via dynamic traffic lights. To validate our approach, we consider a  high-fidelity case study   using the state-of-the-art simulation software package \gls{SUMO}. Preliminary results indicate that \acrshort{DeePC} outperforms existing approaches across various key metrics, including travel time and CO$_2$ emissions,  demonstrating  its potential for effective traffic management.%
\end{abstract}
%\red{General: \begin{itemize}
%    \item Increase the font size of all the figures cause rn they are too small compared to the one in the text.
%    \item Careful w the titles in the literature. Several miss capitalization in the right words
%    \item Never use $\forall$ inline. Online inside an equation.
%\end{itemize}}

\IEEEpeerreviewmaketitle

% \ccmargin{Discuss the difference e between dynamic and static policies?}

\section{Introduction} 
% General intro on traffic
In the ever-growing urban landscapes of our times, the need for efficient and effective traffic congestion management systems has never been more pressing.
%The urgency of this transformation is underscored by a growing body of evidence that paints a bleak picture of the current state of urban traffic control. 
%For instance, 
Recent studies indicate that, on average, commuters in major cities spend approximately 54 hours each year stuck in traffic jams~\cite{schrank2019}.  The economic and environmental toll of traffic congestion is staggering~\cite{fitfor55}.  
As the world continues to urbanize~\cite{un2018}, these figures are projected to increase significantly, unless novel solutions for traffic congestion are developed. 

% The role of data
In the era of digital transformation, data serves as a critical asset in this respect, revealing complex patterns of urban traffic and supporting data-informed  decision-making~\cite{atsi2018}.  
% With the ubiquity of smartphones, the proliferation of \gls{IoT} devices, and the continuous expansion of sensor networks, our society is inundated with a wealth of data~\cite{atsi2018}.   
For instance, machine learning traffic pattern analysis algorithms have demonstrated the potential to enhance traffic flow and alleviate congestion~\cite{ou2018,zhang2021}. 
%However, the implementation of these methodologies raises several concerns, including computational complexity, ease of deployment,  data privacy considerations, and interoperability.
% % This deluge offers the transformative potential to reshape the very foundations of urban traffic management and the unprecedented ability to implement new adaptive urban traffic control strategies.
% Data-driven traffic control
In this context, the emergence of data-driven traffic control algorithms promises to revolutionize how we tackle the  problem   of urban traffic congestion~\cite{li2020}. 
% Goal of the paper
\noindent
With these premises, this paper addresses the urban traffic control problem using the \gls{DeePC} algorithm~\cite{coulson2019data}.  
%Although the algorithm behaves like a classical \gls{MPC} algorithm for Linear Time-Invariant (LTI) systems, its closed-loop performance excels with mildly nonlinear dynamics, making it an effective choice for capturing the complex dynamics of urban traffic and outperforming model-based approaches. 

%% Related work
\textit{Related work:} 
% Traffic
In the past decades, the problem of urban traffic control has attracted the interest of many researchers. Early solutions focus on controlling traffic only in the proximity of traffic lights and intersections~\cite{robertson1991optimizing} and small traffic networks~\cite{papageorgiou2003review}. Over the years results have been developed for larger networks and have often considered as the main objective of the reduction of the total system travel time or the emissions. In \cite{aboudolas2010rolling} the authors model the city as a directed graph and propose a signal control problem based on efficiently solving  a Quadratic Programs (QP).  The authors in \cite{lin2012efficient} propose two different macroscopic models of urban traffic to design an \gls{MPC} able to compute a structured network-wide traffic controller. Control strategies that focus particularly on reducing the emission can be found in~\cite{lin2013integrated,cenedese2022incentive,grontas2023designing,guo2019urban} and references therein. 
% The recent growth of \gls{IoT} within traffic networks and communication networks, together with the adoption of \glspl{CAV} foster researchers into developing novel solutions such as policies influencing drivers' demand~\cite{cenedese2022incentive,grontas2023designing}, multi-vehicle cooperative driving and routing~\cite{guo2019urban}.

% Data-driven control
% 
% The traditional approach to control design has historically favored ``indirect approaches,''  which involve a system identification step followed by model-based control. 
Due to the unprecedented data collection, storage, and computation capabilities, there has been a resurgence of interest in the control community in  ``direct''   control approaches  (see, e.g., the recent survey~\cite{markovsky2021behavioral}). 
Leveraging behavioral system theory~\cite{willems1986timeI}, these
 approaches  infer  optimal decisions directly from observed data. 
% At the heart of this emerging trend lies a seminal result known as the ``fundamental lemma'' due to J. C. Willems and co-authors~\cite{willems2005note}. Leveraging behavioral system theory~\cite{willems1986timeI}, the fundamental lemma establishes that any LTI system can be modelled by a matrix constructed directly from a time series of (sufficiently informative) data.  
A particularly successful direct data-driven algorithm is \gls{DeePC} ~\cite{coulson2019data}, which has been applied in a wide array of practical case studies~\cite{coulson2019data,carlet2020data,huang2019data}. \gls{DeePC} has been applied in the context of traffic control, e.g., to solve control problems related to the coordination of \glspl{CAV}~\cite{wang2023deep}, to vehicle rebalancing in mobility-on-demand systems~\cite{zhu2023data}, and to the dissipation of stop-and-go waves~\cite{wang2023implementation}. However, to the best of our knowledge, \gls{DeePC} has not been applied in the context of urban traffic congestion control due to the large-scale nature of the problem. Motivated and inspired by~\cite{geroliminig:2018:MPC_perimeter_control,gerolimins:2013:optimal_periter_two_regions,geroliminis:2016:snake_MFD}, we propose a tractable formulation by leveraging an aggregate description of the urban traffic dynamics of a city which mitigates the effect of unpredictable events. 

% \apmargin{Changed the format of contributions so as to fit the notation in the first page. Cleaner.}

% For this reason, the concept of \gls{MFD} has been introduced and it manages to grasp in a simple and aggregative form the traffic dynamics in a city's region by collecting data from sensors within it. This concept has been successful in performing perimeter control~\cite{geroliminig:2018:MPC_perimeter_control,gerolimins:2013:optimal_periter_two_regions,geroliminis:2016:snake_MFD} for large metropolis. Moreover, its aggregative nature can mitigate the fluctuation in urban traffic due to unpredictable events. 
% % thanks to description attained as the averaging of many sensors in the region. 
%% Contributions
\textit{Contributions:} 
 The main contributions of the paper are threefold.
 (i) We show that  the \gls{DeePC} algorithm, although fundamentally linear, excels in handling mildly nonlinear systems. 
    Adopting the \gls{DeePC}  algorithm avoids the use of complex nonlinear models, but rather it generates optimal decisions  directly from data. In particular, being able to directly provide control for traffic lights that do not directly regulate the flow among regions is extremely difficult via a traditional \gls{MPC} based approach.
(ii) We show that the building blocks necessary to deploy \gls{DeePC} are simple to create after the partition of the city in homogenous regions, which should be performed only once as in~\cite{geroliminis:2016:snake_MFD}. This allows practitioners to bypass the identification of model parameters and, in case the infrastructure changes, new data can be collected and automatically updated in the algorithm.
    % Nevertheless, \gls{DeePC} seems to capture  and control  well the nonlinear dynamics of the system. Models used in classic algorithms usually require a linearization that can diminish their accuracy. 
    % \item \textit{Hybrid approach:} To ensure that \gls{DeePC} can be implemented in real-time, we have incorporated some knowledge of the traffic network to drastically decrease the control variables. Namely, we predict in a data-driven way the evolution of the density in the regions rather than directly exploiting the raw data without imposing any structure. This also allows us to apply \gls{DeePC} to a system with smoother dynamics and thus, intuitively, with a behavior that can be more easily described via its trajectories.
    (iii) We show that the data-driven approach can outperform   model-based control  in terms of traffic congestion and emission reduction the current state-of-the-art model-based controller via \gls{SUMO} mesoscopic simulations.

\textit{Notation} %We use standard mathematical notation. 
% $\N$ and $\Zge$ denote the set of positive integer numbers and the set of  non-negative integer numbers, respectively. 
% $\R$, $\R^n$ and $\R^{p \times m}$  denote the set of real numbers, the set  of $n$-dimensional vectors with real entries, and the set of $p \times m$-dimensional matrices with real entries, respectively. 
% For every ${p\in\N}$, the set of positive integers $\{1, 2, . . . , p\}$ is denoted by $\mf p$. 
% $I$ denotes the identity matrix.
% $M^{\transpose}$, $\Image M$ and $\ker M$  denote the transpose,  the image and the kernel of the matrix ${M \in \R^{p \times m}}$, respectively. Map, function, and operator are used synonymously. 
% A map $f$ from $X$ to $Y$ is denoted by $f:X \to Y$; $
For every ${p\in\N}$, the set of positive integers $\{1, 2, . . . , p\}$ is denoted by $\mf p$. 
$(Y)^{X}$ denotes the collection of all maps from  $X$ to $Y$.   
for  ${x \in X \cap X^{\prime}}$. 
The restriction of  ${f:X \to Y}$  to a subset  ${X^{\prime} \subset X}$  is denoted by $f|_{X^{\prime}}$ and is defined by $f|_{X^{\prime}}(x)$ 
If $\mathcal{F} \subset (Y)^{X}$, then  $\mathcal{F}|_{X^{\prime}}$ denotes ${\{ f|_{X^{\prime}} \, : \, f \in \mathcal{F}\}}$.  The shift operator $\sigma: {(\R^q)}^\N \to {(\R^q)}^\N$ is defined as $w \mapsto \sigma w$, with ${\sigma w(t) = w(t+1)}$ for all $t\in\N$.  

% \vfill
% %% Organization of the paper
% \textbf{Paper organization}
% The remainder of the paper is organised as follows. 
% Section~\ref{sec:preliminaries} provides basic definitions regarding behavioral systems theory. 
% Section~\ref{sec:methodology} formulates the problem of interest  \bred $\ldots$. \ered
% Section~\ref{sec:simulations} is concerned with \bred stuff. \ered  
% Section~\ref{sec:conclusion} provides a summary of the main results and an outlook to future research directions.   
% %Appendix~\ref{sec:appendix} provides additional background material on \bred stuff. \ered

\section{Preliminaries} \label{sec:preliminaries}

\subsection{Sequences and Hankel matrices}

We consider finite and infinite sequences in ${(\R^q)}^\mathbf{T}$ and ${(\R^q)}^\N$, respectively.  By a convenient abuse of notation, we identify each finite sequence ${w\in(\R^q)}^\mathbf{T}$ with the corresponding vector ${\col(w(1),\ldots,w(T))\in\R^{qT}}$. 
We use the terms sequence and trajectory interchangeably.

The \textit{Hankel matrix} of depth ${L\in\mathbf{T}}$ associated with the finite sequence ${w\in {(\R^q)}^\mathbf{T}}$ is defined as 
\begin{equation} \label{eq:Hankel} 
\! H_{L}(w) \! = \!
\scalebox{0.85}{$
\bma  \nn
\begin{array}{ccccc}
w(1) & w(2)  & \cdots &  w(T-L+1)   \\
w(2) & w(3)  & \cdots &   w(T-L+2)   \\
\vdots  & \vdots  & \ddots & \vdots  \\
w(L) & w(L+1)  & \cdots  & w(T)
\end{array}
\ema
$} . \! \!
\end{equation}

\subsection{Dynamical systems}

A \textit{dynamical system} (or, briefly, \textit{system}) is a triple $\Sigma=(\T,\W,\B),$ where $\T$ is the \textit{time set}, $\W$ is the \textit{signal set}, and $\B \subseteq (\W)^{\T}$ is the \textit{behavior} of the system. 
We exclusively focus on \textit{discrete-time} systems, with ${\T = \N}$ and ${\W = \R^q}$.
%We also routinely identify systems with their behaviors.  

% %\subsubsection{Partitions}
% Given ${m\in\mathbf{q}}$ and a permutation matrix ${\Pi\in\R^{q\times q}}$, the map
% $w\mapsto \Pi w = (u,y)$ defines a \textit{partition} of ${w\in(\R^q)^{\N}}$ into the variables ${u\in(\R^m)^{\N}}$  and ${y\in(\R^{q-m})^{\N}}$.  The map is an \textit{input-output partition} of a system ${\Sigma}$ if $u$ is \textit{free}, \textit{not anticipated}, and \textit{causal}~\cite{willems1989models}, in which case $u$ is the \textit{input} and $y$ is the output of $\Sigma$.  The reader is referred to~\cite[Section 6]{willems1986timeI} and~\cite{willems1989models} for further detail. 

%\subsubsection{Linear, time-invariant, complete systems}
A system $\Sigma$ is \textit{linear} if the corresponding behavior $\B$ is a linear subspace, \textit{time-invariant} if $\B$ is shift-invariant, \textit{i.e.},  if ${w\in \B}$ implies ${\sigma w\in \B}$,  and \textit{complete} if $\B$ is closed in the topology of pointwise convergence~\cite[Proposition 4]{willems1986timeI}.
The model class of all complete  Linear Time-Invariant (LTI)  systems is denoted by $\L^q$.  By a convenient abuse of notation, we write  ${\B \in \L^{q}}$.

%\subsubsection{Structure indices}
The structure of an LTI system is characterized by a set of integer invariants 
known as \textit{structure indices}~\cite{willems1986timeI}. The most important ones are the \textit{number of inputs} $m$, \textit{number of outputs} $p$,  the \textit{lag} $\ell$, and the \textit{order} $n$, see,~\cite[Section 7]{willems1986timeI} for definitions. 
%The structure indices are intrinsic properties of a system,  as they do not depend on  its representation. 
% \apmargin{Should we define integer invariants? Added a short description in case these are needed.} 
% \bred
Every finite-dimensional LTI system  admits a minimal representation   and  can be described by the equations
\beq \label{eq:state-space}
\sigma x = Ax + Bu, \quad y=Cx+Du,
\eeq
% and admits a (\textit{minimal}) \textit{input/state/output representation} 
% \beq \label{eq:partition-ISO} 
% \!  \! \B \!  = \!
% \left\{
% (u,y) \in (\R^{q})^\N \,:\, \exists \,x\in(\R^n)^{\N} \, \textup{s.t.}~\eqref{eq:state-space}~\text{holds}
% \right\},
% \eeq
where
$\scalebox{0.75}{$\bma\!
\begin{array}{cc}
A & B \\
C & D
\end{array}\!\ema $}
 \in \R^{(n+p)\times (n+m)}$ and  $m$, $n$, and $p$ are the number of inputs, the order, and the number of outputs of $\B$, respectively. 
 
%  % The \textit{order} of an LTI system ${\B\in\L^q}$ is the smallest ${n\in\N}$ among all (minimal) state-space representations~\eqref{eq:state-space} and the \textit{lag} is the smallest ${\ell\in\N}$  such  that in a (minimal) state-space representation the \textit{observability matrix} 
% \beq  \nn
% \mathsf{O}_\ell
% =
% \bma
% \begin{array}{cc}
% C \\
% CA \\
% \vdots  \\
% CA^{\ell -1} 
% \end{array} 
% \ema 
% \eeq 
% is  full rank. 

% \ered

\subsection{Data-driven representations of dynamical systems}

Given a  trajectory   ${w_{\tup{d}} \in \R^{qT}}$ of a system ${\B \in \mathcal{L}^{q}}$, it is possible to derive a non-parametric representation of its  finite-horizon  behavior using raw data. We summarize a version of this principle known as the \textit{fundamental lemma}~\cite{willems2005note}.

\begin{lemma}~\cite[Corollary 19]{markovsky2022identifiability} \label{lemma:fundamental_lemma_generalized}
Consider a system ${\B \in \L^{q}}$ with lag ${\ell\in\N}$ and a trajectory of the system ${w_{\tup{d}} \in \B|_{ [1,T] }}$. Assume ${L > \ell}$. Then  
%\beq \label{eq:data-driven-representation} 
$\B|_{ [1,L] } = \Image H_L(w_{\tup{d}})$
%\eeq
if and only if
\beq  \label{eq:generalized_persistency_of_excitation} 
\rank H_L(w_{\tup{d}})   =  mL+ n,
\eeq
where $n$ and $m$ are the order and the number of inputs of the system, respectively.
\end{lemma}

Lemma~\ref{lemma:fundamental_lemma_generalized} is a key result in data-driven control~\cite{markovsky2021behavioral}.  It characterizes all trajectories of given length of an LTI system  in terms of the image of a Hankel matrix.   
%which, in turn, can be constructed directly from raw data. 
This foundational principle can be adapted in various ways, % to suit different assumptions
see the recent survey~\cite{markovsky2021behavioral}. Remarkably, non-parametric representations have found practical use in data-driven control even when dealing with \textit{nonlinear} systems~\cite{berberich2022linear}.

The rank condition~\eqref{eq:generalized_persistency_of_excitation} is known as the  \emph{generalized persistency of excitation} condition~\cite{markovsky2022identifiability}. Note that upper bounds on the structure indices of $\B$ are necessary to check this condition from data.  Alternatively, the rank condition~\eqref{eq:generalized_persistency_of_excitation} can be guaranteed to hold for controllable systems if a certain rank condition on the inputs holds~\cite{willems2005note}.

 % The idea of using non-parametric representations of system is versatile and applicable even to nonlinear systems~\cite{berberich2022linear}. Such foundational principle can be adapted in various ways to suit different assumptions, see the recent survey~\cite{markovsky2021behavioral}. 

% The rank condition~\eqref{eq:generalized_persistency_of_excitation} is known as the \emph{generalized persistency of excitation} condition~\cite{markovsky2020identifiability}. To validate this condition from data, upper bounds on the structure indices of $\B$ are necessary (see Eq.~\eqref{eq:generalized_persistency_of_excitation}). Alternatively, for controllable systems, the rank condition~\eqref{eq:generalized_persistency_of_excitation} holds if certain input rank conditions are satisfied~\cite{willems2005note}.

\section{Methodology} \label{sec:methodology}
% The problem of urban traffic control for a whole city has attracted the interest of the research community in the last decades~\red{\cite{?}}. In order to consider large-scale networks that are able to model a whole city, researchers often rely on aggregate models describing the flow of vehicles among different regions of the cities~\red{\cite{?}}. \apmargin{Add ref?}
% In this paper, we adopt a similar approach. In fact, dividing a city into traffic-wise homogeneous regions allows us to \red{write something on the benefit of an aggregate formulation. Decreasing fluctuations and actionable and scalable solutions.}

\begin{figure}[t]
\centering
\includegraphics[trim={45 0 103 0},clip,width=\columnwidth]{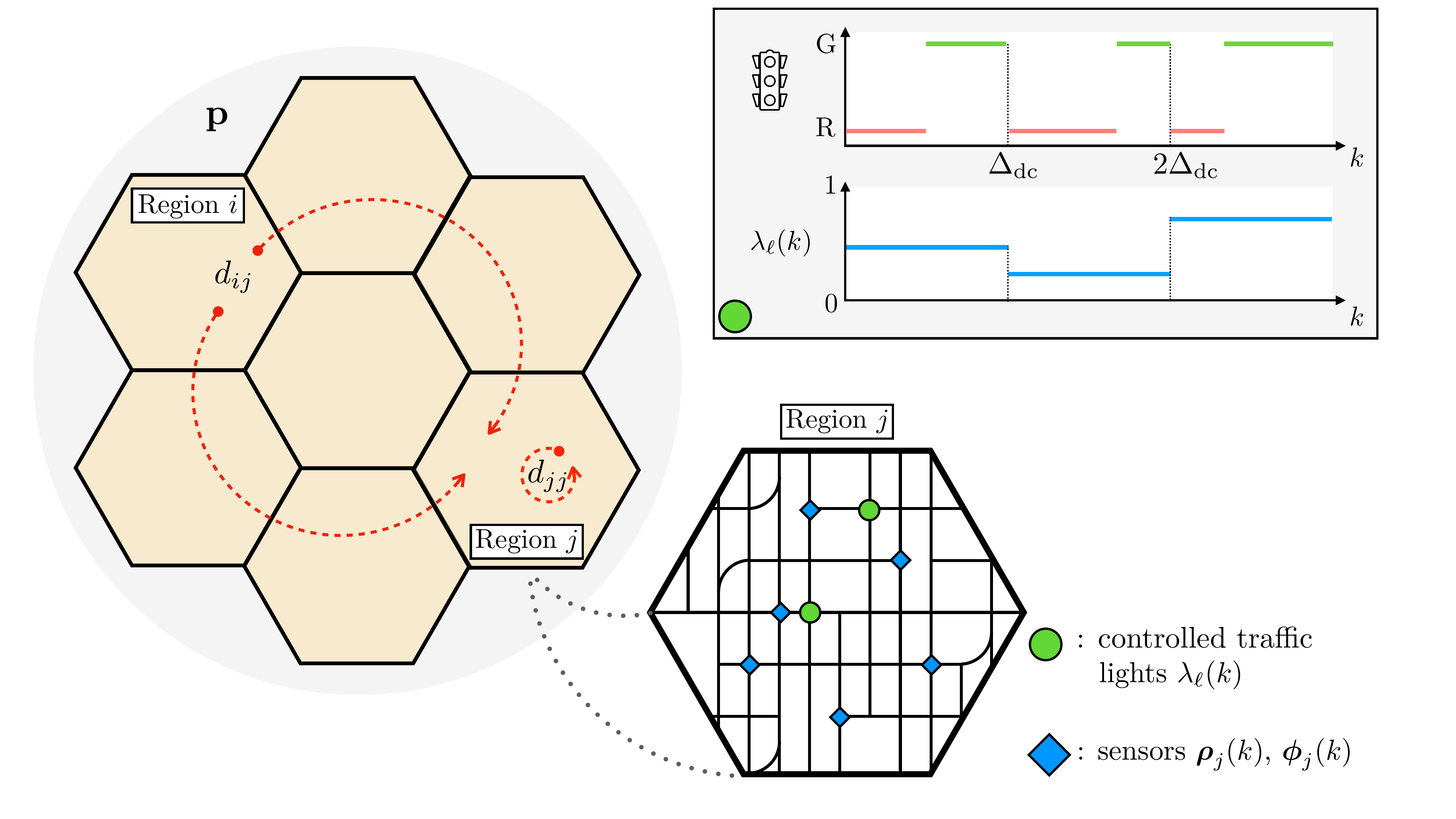}%[width=1\columnwidth} % one column width if needed
\centering
\caption{The city is divided into $\mf p$ regions (the hexagons) based on their \gls{MFD} and the demand $\bs d$ among them. The qualitative detail of how sensors $\mf s_j$ (blue diamonds) and traffic lights $\mf l$ (green circles) might be positioned within a region $j\in\mf p$. In the top part, how the control input $\lambda_\ell$ influences the duty cycle of a traffic light $\ell\in\mf l$.}
\label{fig:info_traffic}
\end{figure}

\subsection{Urban traffic: sensing and dynamics}
\label{ssec:urban_td}
The transportation network of a city is composed of a collection of roads, highways, intersections, and points of interest such as bus stops. As in \cite{geroliminig:2018:MPC_perimeter_control,geroliminis:2016:snake_MFD}, the city can be partitioned in $p\in\N$ regions leveraging the concept of \gls{MFD}. These regions are identified such that the aggregate drivers' behavior traveling through it is as homogeneous as possible. 
% The discussion on the identification of these regions is delayed to  Section \red{?}. %, the interested reader see~\cite{geroliminis:2011:properties_MFD}. 
The set of all regions is denoted by $\mf{p} \coloneqq \{1,\cdots,p\}$.  Modern cities can choose among different types of sensors to assess traffic conditions. Among these, the most common (and reliable) are  \textit{Eulerian} sensors, such as single-loop detectors~\cite{zhanfeng:2001:single_loop_detector}. They can measure the number of vehicles that cross them during a fixed period and the occupancy,
 that is,   the fraction of time during the same period that the sensor detected a vehicle above it \cite{claudel:2008:lagrangian-eulerian_snesing}. Similar data can also be retrieved from other types of sensors, such as \textit{Lagrangian} (or mobile) sensors. 
 This includes vehicles equipped with  transponders and GPS  traveling through the city. 
This allows smaller cities that do not have (or cannot afford) fixed sensors to implement the control scheme designed in the following sections~\cite{claudel:2008:lagrangian-eulerian_snesing}. For simplicity, however, we consider only Eulerian sensors throughout the paper. 
% Merging different data from different types of sensors is not a trivial task~\cite{claudel:2008:lagrangian-eulerian_snesing}. 

% \apmargin{The use of bold for vectors may be inconsistent with the bold notation $\mathbf{n}$}
  We denote the number of sensors in each  region  $i\in\mf p$  as $s_i\in\N$  and the set of all sensors in the region is $\mathbf{s}_i\coloneqq\{1,\cdots,s_i\}$. From each sensor $j\in\mathbf{s}_i$, we can attain, from the occupancy and number of vehicles, the traffic density $\rho_j(t)$\,[veh/km] and flow $\phi_j(t)$\,[veh/h] during the time interval indexed by $t\in\N$ of length $\Delta\in\R_+$. We denote the density and flows measured in region $i\in\mf p$ during $t$  by  
% \apmargin{$k$ or $t$?}
% \apmargin{How about using an equation with actual vectors for clarity?} 
\begin{subequations}
\begin{align}
 \bs \rho_i(t) & \coloneqq \col((\rho_1(t)),\cdots, \rho_{s_i}(t)))\in\R^{s_i}   \\
 \bs \phi_i(t) &\coloneqq \col((\phi_1(t),\cdots,\phi_{s_i}(t)))\in\R^{s_i}\,.
\end{align}
\end{subequations}
% \bred $\bs \rho_i(k) \coloneqq \col((\rho_j(k))_{j\in\mathbf{s}_i})\in\R^{s_i}$ and $\bs \phi_i(k) \coloneqq \col((\phi_j(k))_{j\in\mathbf{s}_i})\in\R^{s_i}$. \ered 
Similarly, the collective vectors of all  densities  and flows in the city during $t$ are, respectively,
\begin{subequations}
\begin{align}
\bs \rho(t) &\coloneqq \col((\bs \rho_i(t))_{i\in\mf p})\in\R^{s} \\
\bs \phi(t) &\coloneqq \col((\bs \phi_i(t))_{i\in\mf p})\in\R^{s}, 
\end{align}
\end{subequations}
 where $s=\sum_{i\in\mf p}s_i$. 
% Shortcomings of the classical models that led us to go to data-driven
%These two variables are deeply intertwined as we will discuss later and provide a good estimation of the traffic conditions in the neighborhood of each sensor. 

The evolution of traffic within the city depends on the number of commuters  entering the system  during every time interval  within each region.  We call this flow of vehicles \textit{demand}. 
% Consider a region $i\in\mf p$, we denote by $\mf n_i\subseteq \mf n$ the set composed of the $n_i$ regions neighboring with $i$.
% and thus there might be a flow of vehicles that travel from one to another. Therefore,
 The flow of vehicles starting their trip in region $i\in\mf p$  at time  $t$ and aiming to end the trip in region ${j\in\mf p}$ is denoted by ${d_{ij}(t)}$\,[veh/h], %represents the flow of vehicles that want to travel from region $i$ to $j$ during the time interval $k$. 
 while $d_{ii}(t)$\,[veh/h] is the \textit{internal trip completion flow}~\cite{geroliminig:2018:MPC_perimeter_control}, which describes the flow of vehicles whose final destination is region $i$. 
The collective vector of demands associated with region $i$ is denoted by $\bs d_i(t) \coloneqq \col((d_{ij}(t))_{j\in\mf p})\in\R^{p}$ while the vector of all the demands among all regions is $\bs d(t) \coloneqq \col((\bs d_{i}(t))_{i\in\mf p})\in\R^{p^2}$. Hereafter, we consider the \textit{nominal} value of $\bs d(t)$ as an exogenous and known quantity. How to estimate the demand among regions is a well-known problem that has been extensively studied~\cite{foulds:2013:origin_destination_matrix,olmos:2018:urban_traffic}.
% , and these flows are usually described via an origin-destination matrix. Nevertheless, the demand is greatly affected by the uncertain environment in which personal choices, weather conditions, and period of the year can highly affect the accuracy of the estimation, and thus $\bs d(t)$ is highly affected by noise. 
% \ccmargin{Only the demand spawn in the region of a certain time is known and where they want to go then they are "lost" in the dynamics of the regions that are somehow hidden within the densities of the regions.}

% 
% \smallskip
% \begin{remark}[Routing]
%     We do not need it this is quite a pro compared to other things.\hfill$\QEDopen$
% \end{remark}
% \smallskip
% 

Policymakers throughout the years have designed a plethora of interventions to reduce urban traffic 
 congestion~\cite{elokda2022carma}. 
% \apmargin{Removed in the previous Section}
%\strikt , as discussed in the previous \red{Section}. 
Among dynamic policies, there is the installation of controllable traffic  lights~\cite{ishu:2022:traffic_signal_control}.  By strategically controlling the flow of vehicles traveling through   an intersection,   it is possible to indirectly influence  $\bs \phi_i(t)$ and $\bs \rho_i(t)$ in each region.  Given  
 two regions ${i,j\in\mf p}$, if one (or multiple)  intersection  used by drivers to move  from $i$ to $j$  allows for a smaller flow to pass through, then the vehicles will accumulate increasing the density in  $i$  and decreasing the flow from $i$ to $j$. 
%If one can control all the access points from region $i$ to $j$ then $d_{ij}(t)$ can be directly controlled.
The flow of vehicles that can travel through a intersection can be controlled by varying the ratio  green  $\lambda(t)$  of the duty cycle $\Delta_{\tup{dc}}\in\N$, see Fig.~\ref{fig:info_traffic}. 
% The subscript ``tl'' highlights that this control is directly applied to the traffic lights in the city. \apmargin{Small notation issue: $\ell$}
% For notational convenience, we assume that $\Delta_{\tup{dc}}$ is a multiple of $\Delta$.
Therefore, the control input $\lambda_{\ell}(t)\in[\underline{\lambda}_\ell,\overline{\lambda}_\ell]\subseteq[0,1]$ where $\lambda_{\ell}(t)=1 $ means that the traffic light is  always green and  $\lambda_{\ell}(t)=0$ red for the whole duty cycle.  
The vector of all  input associated with the $l$ controlled traffic lights is denoted by 
 $\bs \lambda(t) = \col((\lambda_{\ell} (t))_{\ell\in\mf l})$, where $\mf l = \{1,\cdots,l\}$.  
%\ccmargin{In the dynamics part, discuss the fact that we might want to add the flows in order to "help" the linearization of the model.}
%\apmargin{How about calling the traffic light $\lambda$? And then just having $u=(\lambda,d)$?}

We  model   the nonlinear dynamics associated with the density evolution measured by the sensors as
\begin{equation}
    \label{eq:density_dyn}
    {\bs \rho}(k+1) = f({\bs \rho}(t),\bs \lambda(t),\bs d(t)),
\end{equation}
 where $f:\R^{s+l+p^2}\rightarrow\R^{s}$ is not known. 
 % As done in \cite{geroliminig:2018:MPC_perimeter_control}, 
 % Thanks to the tight relation between flow and density, discussed more in detail in Section~\ref{sec:DeePC_traffic}, we can focus hereafter only on the density dynamics since this will allow us to reduce the number of variables necessary to compute the optimal control action. 
 % the strong connection between flow and density, which will be explicitly described in Section~\ref{sec:DeePC_traffic},  to reduce the number of variables that are necessary to compute the control action. 
 This  choice  mitigates the course of dimensionality associated with the proposed data-driven approach. Moreover,  as demonstrated by our numerical investigations,   the density usually has smoother trajectories that are easier to predict via a data-driven approach.
 % Therefore, hereafter we focus only on the density dynamics. 
 
 As discussed in \cite{geroliminis:2011:properties_MFD}, aggregating the data from the sensor within a region helps reduce the variability and capture the macroscopic variation in  density. 
 % within such a region minimizing the effect of outliers measured by the sensors in case disturbances due to accidents, unusual humans behaviors or sensor fault. There are different types of aggregating functions, for example, one can choose a weighted mean depending on the accuracy of the considered sensor. In general, 
 An  aggregating function has the role of mapping the sensors' measured densities to the associated region density. 
 So, the \textit{average} density of region $i$ is attained as $\overline \rho_i(t) = h_i(\bs \rho_i(t))$, where $h_i:\R^{s_i}\rightarrow\R$ is the aggregating function of region $i\in\mf p$.
 Consequently, we  define  the complete aggregation function as $h:\R^{s}\rightarrow\R^p$ and
 \begin{equation}
     \label{eq:aggregation_fcn}
     \overline {\bs\rho}(t) =  h(\bs\rho(t))\coloneqq\col\big((h_i({\bs \rho_i}(t)))_{i\in\mf p}  \big)\,.
 \end{equation}
Using the density dynamics of the city regions in \eqref{eq:density_dyn}--\eqref{eq:aggregation_fcn} and the definition in Section~\ref{sec:preliminaries}, we  now  introduce the behavior $\mc B_{\tup c}$ that describes the associated dynamical system  
% \apmargin{Why $\mc B_{\tup c}$?}
\begin{equation} 
    \label{eq:behavior_city}
\mc B_{\tup c} = \{(\bs \lambda, \bs d, \overline {\bs\rho})\in\R^{m+p} \,:\, \exists {\bs\rho}\in\R^s \,\text{s.t \eqref{eq:density_dyn}--\eqref{eq:aggregation_fcn} hold}  \}    , 
\end{equation}
where $m=l+p^2$ and $u=(\bs \lambda, \bs d)\in\R^m$ is the complete input, divided in exogenous and controllable, and $\overline {\bs\rho}$ the output.

% In the next section, we introduce a theoretical formulation of the data-driven algorithm used to control the dynamical system defined above.

\subsection{The \gls{DeePC} algorithm} \label{ssec:DeePC}

% The \gls{DeePC} algorithm relies on the fundamental lemma to perform predictive control directly from data~\cite{coulson2019data}. We provide a brief overview of the setup and assumptions for completeness.

\subsubsection{Setup and assumptions}
Consider a (possibly unknown) LTI system ${\B\in\L^{m+p}}$, with $m$ inputs and $p$ outputs.  Assume that data recorded offline from system $\B$  are  available. Specifically, assume that an input sequence $u_{\tup{d}} =\col(u_{\tup{d}}(1),\cdots,u_{\tup{d}}(T)) \in \R^{mT}$ of given length $T\in \N$ is applied to the system $\B$ and that the corresponding output sequence is $y_{\tup{d}}=\col(y_{\tup{d}}(1),\cdots,y_{\tup{d}}(T))\in \R^{pT}$.  The subscript ``$\textup{d}$'' is used to indicate that these are sequences of data samples collected  offline.  
% \ccmargin{We never introduced $\mathbf{T}$.}
Finally, 
  let $\Tini\in \N$ and $\Tf\in \N$, with $\Tini + \Tf \le T,$ and  
assume that the sequence $w_{\tup{d}}\in \R^{(m+p)T}$, defined as $w_{\tup{d}}(t) = \col(u_{\tup{d}}(t),y_{\tup{d}}(t))$ for ${t\in \mathbf{T}}$,
 satisfies the generalized persistency of excitation condition~\eqref{eq:generalized_persistency_of_excitation},  where  ${L=\Tini+\Tf}$ and $n$ is the order (or an upper bound on the order) of the system. 

\subsubsection{Data organization}\label{sec:data_org}
Next, partition the input/output data into two parts which we call \emph{past data} and \emph{future data}. Formally,  given the time horizons $\Tini \in \N$ and $\Tf \in \N$ associated with the past data and the future data, define 
\begin{equation}\label{eq:UpUfYpYf}
\begin{pmatrix}
\Up \\ \Uf 
\end{pmatrix}= H_{\Tini+\Tf}(u_{\textup{d}}), \quad
\begin{pmatrix}
\Yp \\ \Yf 
\end{pmatrix}= H_{\Tini+\Tf}(y_{\textup{d}}),
\end{equation}
where ${\Up\in\R^{(m\Tini)\times (T-\Tini+1)}}$ consists of the first $\Tini$ block-rows of the matrix $H_{\Tini+\Tf}(u_{\textup{d}})$ and ${\Uf \in\R^{(m\Tf)\times (T-\Tf+1)}}$ consists of the last $\Tf$ block-rows of the matrix $H_{\Tini+\Tf}(u_{\textup{d}})$ (similarly for $\Yp$ and $\Yf$), respectively. In the sequel, past data denoted by the subscript ``$\mathrm{p}$''  is used to estimate the initial condition of the underlying state, whereas the future data denoted by the subscript ``$\mathrm{f}$''  is used to predict the future trajectories. 

\subsubsection{State estimation and trajectory prediction}
By the fundamental lemma, any trajectory of the  finite-horizon  behavior $\B|_{ [1,\Tini+\Tf] }$ of length $\Tini+\Tf$ can be constructed using the data collected offline.  Under the assumptions of Lemma~\ref{lemma:fundamental_lemma_generalized},  a trajectory $\col(\uini,\yini,u,y)$ belongs to $\B|_{ [1,\Tini+\Tf] }$ if and only if there exists $g\in \R^{T-\Tini-\Tf+1}$ such that
\begin{equation}\label{eq:datamodel}
\begin{pmatrix}
\Up \\ \Yp \\ \Uf \\ \Yf 
\end{pmatrix}
g=
\begin{pmatrix}
\uini \\ \yini \\ u \\ y
\end{pmatrix}.
\end{equation}
For ${\Tini \geq \ell}$, the lag of the system,  the  output $y$ is uniquely determined~\cite{markovsky2021behavioral}. Intuitively, the trajectory $\col(\uini,\yini)$ specifies the underlying initial state from which the trajectory $\col(u,y)$ evolves. This allows one to predict future trajectories based on a given initial trajectory $\col(\uini,\yini)\in\B|_{ [1,\Tini] }$, and the precollected data in $\Up$, $\Uf$, $\Yp$, and $\Yf$. Indeed, given an initial trajectory  $\col(\uini,\yini)\in\B|_{ [1,\Tini] }$  of length ${\Tini \geq \ell}$ and a sequence of future inputs $u\in \R^{m \Tf }$, the first three block equations of~\eqref{eq:datamodel} can be solved for $g$. 
%\ccmargin{Shall we use $\ell$ instead of $l$?}
The sequence of future outputs is then given by $y=\Yf g$. Conversely, given a desired reference output $y$ an associated control input $u$ can be calculated.

\subsubsection{\gls{DeePC} algorithm}
Given the future time horizon $\Tf \in \N$, a reference trajectory for the output $\hat y=(\hat y_{0},\hat y_{1},\cdots)\in (\R^{p})^{\N}$ and input $\hat u=(\hat u_{0},\hat u_{1},\cdots)\in (\R^{m})^{\N}$, past input/output data $
\wini = \col(\uini,\yini)\in \B|_{ [1,\Tini] }$, input constraint set $\mathcal{U}\subseteq \R^{mT_{\tup f}}$, output constraint set $\mathcal{Y}\subseteq \R^{pT_{\tup f}}$, output cost matrix $Q\in \R^{p\times p}$, and control cost matrix $R \in \R^{m\times m}$,  and 
regularization function $\psi:\R^{T-L+1}\to \R$, 
the \gls{DeePC} algorithm relies on the solution of the following optimization problem:
\begin{align}\label{eq:DeePC}
\underset{u,y,g}{\text{min}}\,
&  
\displaystyle
\sum_{k=1}^{\Tf}\left\|y(k)-\hat{y}(t+k)\right\|_Q^2 
+\left\|u(k)-\hat{u}(t+k)\right\|_R^2 +\psi(g)  \nonumber \\
\text{s.t.\,}
& \begin{pmatrix}
\Up \\ \Yp \\ \Uf \\ \Yf
\end{pmatrix}g
=\begin{pmatrix}
\uini \\ \yini \\ u \\ y
\end{pmatrix}, \\
& u\in \mathcal{U}, \: y\in\mathcal{Y}.\nonumber
\end{align}%

%We are now ready to present the \gls{DeePC} algorithm.
\setlength{\intextsep}{10pt}
\begin{algorithm}[h!]
	\caption{DeePC}
	\label{alg:robustdeepc}
	\textbf{Input:} 
	Data trajectories $ w_{\tup{d}}  =\col(u_{\tup{d}},y_{\tup{d}})\in\R^{(m+p)T}$, 
    most recent input/output measurements $\wini =\col(\uini,\yini)\in\R^{(m+p)\Tini}$, 
    a reference trajectories $\hat y=(\hat y_{0},\hat y_{1},\cdots)\in (\R^{p})^{\N}$, $\hat u=(\hat u_{0},\hat u_{1},\cdots)\in (\R^{m})^{\N}$,
    input constraint set ${\mathcal{U}\subseteq \R^{mT_{\tup f}}}$, output constraint set ${\mathcal{Y}\subseteq \R^{pT_{\tup f}}}$, output cost matrix   ${Q\in \R^{p\times p}}$, and control cost matrix ${R \in \R^{m\times m}}$.
	\begin{algorithmic}[1]
		\STATE \label{step:deepcbegin} Solve~\eqref{eq:DeePC} for ${g}^{\star}$.
		\STATE Compute optimal input sequence $u^{\star} = \Uf {g}^{\star}$.
		\STATE Apply optimal input sequence $(u_t,\cdots,u_{t+j-1})=(u_1^{\star},\cdots,u_{j}^{\star})$ for some $j\leq \Tf$.
		\STATE Set $t$ to $t+j$ and update ${u}_{\textup{ini}}$ and ${y}_{\textup{ini}}$ to the $\Tini$ most recent input/output measurements.
		\STATE Return to~\ref{step:deepcbegin}.
	\end{algorithmic}
\end{algorithm}%

\begin{remark}
As demonstrated by the numerical investigation in Section~\ref{sec:simulations}, our formulation leads to satisfactory performance. However, extra slack variables can be used to account for the presence noise explicitly~\cite{coulson2019data}. A thorough exploration of this issue is deferred to a future publication.
\end{remark}

\subsection{\gls{DeePC} for urban traffic control}\label{sec:DeePC_traffic}

We are now ready to  introduce  \gls{DeePC} for the dynamical system 
$\mc B_{\tup c}$ describing the evolution of urban traffic. 
\subsubsection{Behavioral representation and constraints}
The data matrix on the left-hand side in \eqref{eq:datamodel} can be constructed   using  historical data. For $t\in\bf T$ the collection of input and output data used are respectively defined  as
\begin{align*}
    u_{\tup d }(t) &\coloneqq  \col(\bs \lambda(t),\, \bs d(t))\\
    y_{\tup d }(t) & \coloneqq  \col(\overline{\bs \rho}(t))\,.
\end{align*}
Using these sequences, we can construct the matrices in \eqref{eq:UpUfYpYf} as discussed in Section~\ref{sec:data_org}.  The  data are collected offline and thus can be chosen among many different traffic scenarios that excited the dynamical system $\mc B_{\tup c}$ in different ways capturing a wide variety of the behaviors of the system. To construct $\uini$ and $\yini$ for the current time interval $t\in\N$,  we %can 
simply collect the previous $\Tini$ values of the input and outputs,
\begin{align*}
    \uini &\coloneqq \col(\bs \lambda(t-\Tini),\, \bs d(t-\Tini),\cdots, \bs \lambda(t),\, \bs d(t) ), \\
    \yini &\coloneqq \col(\overline{\bs \rho}(t-\Tini),\cdots, \overline{\bs \rho}(t))\,.
\end{align*}

% \apmargin{Just added some commas}
The control input
% $\bs \lambda$, that is a component of $u$, 
has to satisfy the box constraints  
$\bs \lambda(k)\in[\underline{\bs \lambda},\overline{\bs \lambda}]\subseteq \R^{l}$, where $\underline{\bs \lambda}=\col(\underline{\lambda}_1,\cdots , \underline{\lambda}_{l})$ and $\overline{\bs \lambda}$  is defined similarly. Moreover,  the green ratio cannot be changed during a duty cycle but only at the end of it, hence every $\Delta_{\tup{dc}}$ time intervals of length $\Delta$.  
% the length of the time interval is $\Delta$ (or a multiple of it), while due to the duty cycle of the traffic lights the input must remain constant for at least $\Delta_{\tup{dc}}$ time intervals of length $\Delta$. 
This translates into the following linear constraint on the variable $\bs u$
$$Mu=\bs 0\,,$$ 
% \apmargin{Again, maybe change $\bf l$? }
where $M\in\R^{mT_{\tup f}\times mT_{\tup f}}$ is used to impose for every $\ell\in \bf l   $ and $k\leq T_{\tup f}$ multiple of $\Delta_{\tup{dc}}$ the constraint 
$$\lambda_{\ell}(k+1)=\lambda_{\ell}(k+2)=\cdots=\lambda_{\ell}(k+\Delta_{\tup{dc}})\,.$$
On the contrary, the demand $\bs d$ is assumed to be a known exogenous input. We impose it to be fixed and equal to the known demand $\overline{\bs d}=\col((\bs d(t),\cdots,\bs d(t+T_{\tup f})))\in\R^{p^2T_{\tup f}}$ over the prediction horizon, i.e., $Du=\overline {\bs d}$ with $D\in\R^{mT_{\tup f}\times p^2T_{\tup f}}$ being a suitable matrix. Therefore, the set of constraints for the inputs over the whole prediction horizon reads as
\begin{equation}\label{eq:cal_U}
    \mc U=\left\{u\in \big([\underline{\bs \lambda},\overline{\bs \lambda}]\times\R_+^{p^2}\big)^{T_{\tup f}} \,|\, Mu=0,\, Du=\overline {\bs d}\right\}.
\end{equation}

The  box  constraints on the output ensure  that the density in the region remains below the \textit{grid-lock density} $\overline{\bs \rho}_{\max}\in\R_+^p$, hence $y(t)\in[0,\overline{\bs \rho}_{\max}]$ and \mbox{$\mc Y\coloneqq[0,\,\overline{\bs \rho}_{\max}]^{T_{\tup f}} $}.

Finally, in the third step of the  \gls{DeePC} algorithm, we apply to the system the first part of the computed optimal input $\bs \lambda^\star$. In this application of urban traffic control, we use $j=\Delta_{\tup{dc}}$ since the control input $\bs \lambda$ cannot be changed during that interval.
% , and thus it would be inefficient to apply a shorter sequence.

% \ccmargin{discuss how to handle the presence of $d(t)$ in the input and whether we need it in the control action. Shall we reformulate the DeePC to address since the beginning the presence of exogenous inputs? Or discuss this possibility in the previous section?}

\begin{figure}[t]
\centering
\includegraphics[width=\columnwidth]{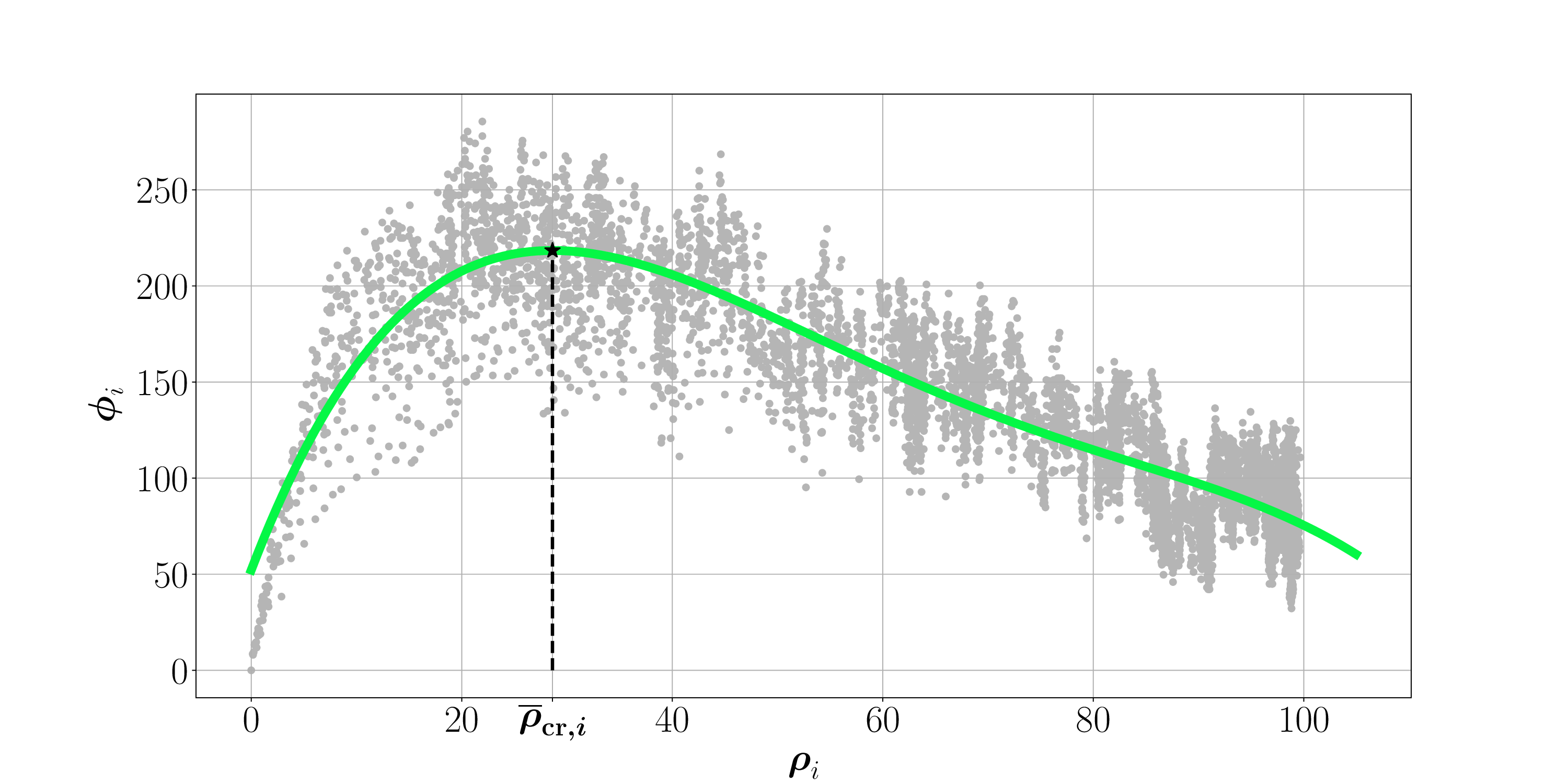}
\centering
\caption{The \gls{MFD} for the region $i\in \mf p$, the historic measured data $\bs{\rho_i},\bs{\phi_i}\in\R^{s_i}$ by the sensors $\mf{s_i}$ (light gray dots). The associated \gls{MFD} for region $i\in \mf{p}$ (solid green line) displaying the relation between $\overline{\bs\rho}_i$ and $\overline{\bs\phi}_i$. 
The critical density $\overline{\bs{\rho}}_{\tup{cr},i}$ is the density maximizing the flow in the region $\overline{\bs\phi}_i$,  while $\overline{\bs\rho}_{\max}$ is the density at which the flow is zero.}
\label{fig:MFD}
\end{figure}

% \apmargin{Isn't it strange that MFD appears in the Figure first?}

\subsubsection{Reference trajectory and cost function}
% The final component missing to cast the optimization problem in \eqref{eq:DeePC} for this particular application are the reference signals $\hat y$ and $\hat u$. The optimal operational point is the one that maximizes the output of the traffic network, hence allowing the maximum flow of vehicles among regions, making drivers complete their trips in the fastest way possible. 
Motivated and inspired by~\cite{gerolimins:2013:optimal_periter_two_regions},  % As done in
we take advantage of the concept of \gls{MFD} to estimate what is the optimal density for each region, $\hat y$. The concept of \gls{MFD} formalizes the relation between the flow $\bs \phi_i$ and density $\bs\rho_i$ within a region $i\in\bf p$. As shown in Fig.~\ref{fig:MFD}, we can use the \gls{MFD} to identify each region's $\overline\rho_{\tup{cr},i}$ that corresponds to the density ensuring the maximum flow within the region, thus $\overline{\bs \rho}_{\tup{cr}}=\col(( \overline\rho_{\tup{cr},i})_{i\in\bf p})$ and $\hat y=(\overline{\bs \rho}_{\tup{cr}},\overline{\bs \rho}_{\tup{cr}},\cdots)$. In \eqref{eq:DeePC}, we minimize the distance between the regions' densities and the critical ones. In a congested scenario, this leads to also maximizing the flow within the region.
The value of  $\overline{\bs \rho}_{\tup{cr}}$ can be initially obtained via historical data, yet while running the algorithm the closed-loop behavior of the system may vary and the newly collected data can be used to recompute the \gls{MFD} and update $\hat y$ if necessary. Traffic evolution throughout a single day has two major peaks during the morning and evening commute. To achieve better performance, practitioners can select a time-varying $\overline{\bs \rho}_{\tup{cr}}$. These changes would not affect the theoretical formulation proposed and are relatively straightforward to implement. 

The reference value of the inputs $\hat u$ is composed of the one associated with $\bs \lambda$, and the one for the exogenous output. The latter is exogenous and set equal to $\bs{d}$ according  to \eqref{eq:cal_U}.
The former represents the normal operation mode of the traffic lights and can be chosen constant if the amount of time traffic lights are green during a duty cycle does not vary throughout the day. % \ccmargin{what about $d$, also as is written now we are weighting also the $d$.}

 In the cost function in \eqref{eq:DeePC}, the matrices $R$ and $Q$ are chosen positive definite. Notice that the components of $\lVert u - \hat u\rVert_R$ associated with $\bs d$ are always equal to zero due to the chosen reference for it and the constraints on $u$.
% to not weight the components of $u$ associated with the demand. In particular, $R$ has a block diagonal structure where the blocks associated with $\bs d$ are composed of all zeros. To ensure a fast convergence of the optimization problem we choose $Q$ positive definite. 
The regularization term $\psi$ has been chosen as in \cite{coulson2019data}. 
% \ccmargin{Add citation on regularization.}
% Notice that in general the optimization problem only requires mere convexity of the cost function.  

% This concludes the formulation of \gls{DeePC} for urban traffic control since we have defined all the quantities necessary to cast \eqref{eq:DeePC} and solve via Algorithm~\ref{alg:robustdeepc}.

\section{Simulations} \label{sec:simulations}

  We illustrate the performance of our urban traffic congestion control framework with two case studies: a lattice network and a preliminary study of the traffic network of Z\"urich.  We have tested the \acrshort{DeePC} algorithm against a model-based predictive control (\acrshort{MPC}) formulation devised in \cite{linearMPC}. We studied the behavior of \acrshort{DeePC} and \acrshort{MPC} in two scenarios: a congested and an uncongested scenario. The main difference between the two is the demand experienced by the network. Both traffic networks are modeled using the open-source state-of-the-art simulation software package \gls{SUMO}~\cite{sumo} which offers the possibility of simulating multi-model large-scale traffic networks both in a microscopic and mesoscopic fashion and provides estimates on the emissions produced.  All simulations are conducted on a 3.6 GHz AMD Rizen 7 4000 Series processor. 

%\apmargin{Removed ``advertisement'' of the GitLab repo for now. Better advertise it later once it's fully polished and complete.}
%and the code is publicly available at:~\url{https://gitlab.ethz.ch/apadoan/fix-the-traffic-in-zurich}.

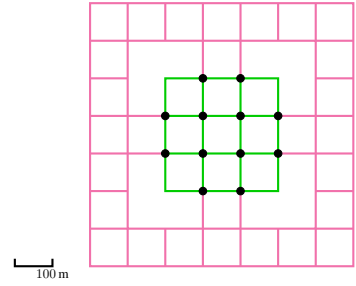
\begin{figure}[t]
\centering
%\includegraphics[scale = 0.4]{./imgs3/regions_actuators.pdf}%[width=1\columnwidth} % one column width if needed
%\begin{tikzpicture}[scale=0.75, every node/.style={scale=0.75}]
\begin{tikzpicture}[scale=1, every node/.style={scale=1}]

% Anchor points
\node at (-6.5,5) {};
\node at (-1.5,5.5) {};
\node at (-1.5,1) {};
\node at (-6.5,1) {};

% Outerr ectangle
%% Outer rectangular region
\draw[thick, color = magenta!70!white]   (-5.5,5) rectangle (-2,1.5);
\draw [thick, color = magenta!70!white] (-5,4.5) rectangle (-2.5,2);
% Top 
\draw [thick, color = magenta!70!white](-5,5) -- (-5,4.5);
\draw [thick, color = magenta!70!white](-4.5,5) -- (-4.5,4.5);
\draw [thick, color = magenta!70!white](-4,5) -- (-4,4.5);
\draw [thick, color = magenta!70!white](-3.5,5) -- (-3.5,4.5);
\draw [thick, color = magenta!70!white](-3,5) -- (-3,4.5);
\draw [thick, color = magenta!70!white](-2.5,5) -- (-2.5,4.5);
% Bottom 
\draw [thick, color = magenta!70!white](-5,2) -- (-5,1.5);
\draw [thick, color = magenta!70!white](-4.5,2) -- (-4.5,1.5);
\draw [thick, color = magenta!70!white](-4,2) -- (-4,1.5);
\draw [thick, color = magenta!70!white](-3.5,2) -- (-3.5,1.5);
\draw [thick, color = magenta!70!white](-3,2) -- (-3,1.5);
\draw [thick, color = magenta!70!white](-2.5,2) -- (-2.5,1.5);
% Left 
\draw [thick, color = magenta!70!white](-5.5,4.5) -- (-5,4.5);
\draw [thick, color = magenta!70!white](-5.5,4) -- (-5,4);
\draw [thick, color = magenta!70!white](-5.5,3.5) -- (-5,3.5);
\draw [thick, color = magenta!70!white](-5.5,3) -- (-5,3);
\draw [thick, color = magenta!70!white](-5.5,2.5) -- (-5,2.5);
\draw [thick, color = magenta!70!white](-5,2) -- (-5.5,2);
% Right 
\draw [thick, color = magenta!70!white](-2.5,4.5) -- (-2,4.5);
\draw [thick, color = magenta!70!white](-2.5,4) -- (-2,4);
\draw [thick, color = magenta!70!white](-2.5,3.5) -- (-2,3.5);
\draw [thick, color = magenta!70!white](-2.5,3) -- (-2,3);
\draw [thick, color = magenta!70!white](-2.5,2.5) -- (-2,2.5);
\draw [thick, color = magenta!70!white](-2,2) -- (-2.5,2);

%% Connections
% Top
\draw [thick, color = magenta!70!white](-4,4) -- (-4,4.5);
\draw [thick, color = magenta!70!white](-3.5,4) -- (-3.5,4.5);
% Bottom
\draw [thick, color = magenta!70!white](-4,2) -- (-4,2.5);
\draw [thick, color = magenta!70!white](-3.5,2) -- (-3.5,2.5);
% Left
\draw [thick, color = magenta!70!white](-5,3.5) -- (-4.5,3.5);
\draw [thick, color = magenta!70!white](-4.5,3) -- (-5,3);
% Right
\draw [thick, color = magenta!70!white](-3,3.5) -- (-2.5,3.5);
\draw [thick, color = magenta!70!white](-2.5,3) -- (-3,3);

%% Inner rectangular region
\draw [thick, color = green!80!black] (-4.5,4) rectangle (-3,2.5);
% Horizontal
\draw [thick, color = green!80!black](-4.5,3.5) -- (-3,3.5);
\draw [thick, color = green!80!black](-4.5,3) -- (-3,3);
% Vertical
\draw [thick, color = green!80!black](-4,4) -- (-4,2.5);
\draw [thick, color = green!80!black](-3.5,4) -- (-3.5,2.5); 

%% Circles
\node[draw, shape = circle, fill=black, minimum size = 0.1 cm, inner sep =0pt] at (-4,3.5) {};
\node[draw, shape = circle, fill=black, minimum size = 0.1 cm, inner sep =0pt] at (-3.5,3.5) {};
\node[draw, shape = circle, fill=black, minimum size = 0.1 cm, inner sep =0pt] at (-4.5,3.5) {};
\node[draw, shape = circle, fill=black, minimum size = 0.1 cm, inner sep =0pt] at (-3,3.5) {};
\node[draw, shape = circle, fill=black, minimum size = 0.1 cm, inner sep =0pt] at (-4,3) {};
\node[draw, shape = circle, fill=black, minimum size = 0.1 cm, inner sep =0pt] at (-3.5,3) {};
\node[draw, shape = circle, fill=black, minimum size = 0.1 cm, inner sep =0pt] at (-4.5,3) {};
\node[draw, shape = circle, fill=black, minimum size = 0.1 cm, inner sep =0pt] at (-3,3) {};
\node[draw, shape = circle, fill=black, minimum size = 0.1 cm, inner sep =0pt] at (-3.5,4) {};
\node[draw, shape = circle, fill=black, minimum size = 0.1 cm, inner sep =0pt] at (-4,2.5) {};
\node[draw, shape = circle, fill=black, minimum size = 0.1 cm, inner sep =0pt] at (-4,4) {};
\node[draw, shape = circle, fill=black, minimum size = 0.1 cm, inner sep =0pt] at (-3.5,2.5) {};
% Scale
\draw [thick](-6.5,1.5998) -- (-6.5,1.5) -- (-6,1.5) -- (-6,1.5998);
\node at (-6,1.400) {\scalebox{0.5}{$100$\,m}};
%\node at (-6.5,1.400) {\scalebox{0.5}{$0$}};
\end{tikzpicture}
\centering
\caption{Lattice traffic network.
The network is partitioned into an outer region  (magenta) and in inner region (green). Black dots represent the controlled traffic lights.}
\label{fig:net}
\end{figure}

\subsection{Case study 1: lattice network}

Consider a lattice network whose topology is shown in Fig.~\ref{fig:net}.
The network is composed of ${208}$ roads (each one equipped with a sensor) connected by ${64}$ intersections, of which ${12}$ are controlled and act as actuators, represented by black dots in Fig.~\ref{fig:net}. All  vehicles are assumed to be cars. The simulation covers a period of one hour. 
 Perimeter control requires the network to be partitioned into regions, composed of roads that are similar in terms of average density. This allows for the production of a low scatter \acrshort{MFD}. 
 Fig.~\ref{fig:net} shows the partitioning used, where we define two regions the outer one (magenta) and the inner one (green).

% We tested \acrshort{DeePC} against a model based predictive control (\acrshort{MPC}) formulation devised in \cite{linearMPC}.
% We studied the behaviour of \acrshort{DeePC} and \acrshort{MPC} in two scenarios: a congested and an uncongested scenario.
% The main difference between the congested and uncongested scenarios is the demand that the network experiences, 
The demand in the network has a clear direction, all trips generated move from the outer region to the inner region. This simulates the flow of vehicles traveling during the morning peak hours from the outside of a city to the city center.
% that the work center of a city experiences during the morning rush hours. 
The demand has a triangular shape, peaking at minute 30 with a base of length 800 [s]. The maximum flow is 180 [veh/h] in the uncongested setting and 1080 [veh/h] in the congested scenario.
% \begin{figure}[h!]
% \centering
% \includegraphics[scale = 0.2]{./imgs3/demands.pdf}%[width=1\columnwidth} % one column width if needed
% \centering
% \caption{The demands of the two scenarios}
% \label{fig:demands}
% \end{figure}
% Fig.~\ref{fig:demands} shows the demand associated with each scenario. 
 The average travel time without control in the uncongested scenario is 9.3 minutes and 6.7 minutes of waiting time due to traffic congestion. In the congested case, the average travel time is 20.6 minutes with 18.4 minutes of waiting time.

\begin{table}[htb]
\caption{Uncongested scenario}\label{tab:uncongested}
    \centering
    \begin{tabular}{@{}llll@{}}
        \toprule
        {} & DeePC & MPC & No Control \\
        \midrule
        Travel Time (min) & $\boldsymbol{5.84}$ & 6.56 & 9.33\\
        Waiting Time (min) & $\boldsymbol{2.89}$ & 3.51 & 6.74\\
        CO Emissions(kg) & $\boldsymbol{0.039}$ & 0.046 & 0.076\\
        CO$_2$ Emissions (kg) & $\boldsymbol{0.958}$ & 1.075 & 1.501 \\
        HC Emissions ($10^{-4}$kg) & $\boldsymbol{2.03}$ & 2.38 & 3.84 \\
        PMx Emissions ($10^{-4}$kg) & $\boldsymbol{0.20}$ & 0.23 & 0.34 \\
        NOx Emissions ($10^{-4}$kg) & $\boldsymbol{4.16}$ & 4.69 & 6.68 \\
        \bottomrule
    \end{tabular}
\end{table}
\begin{table}[htb]%{Table II: Congested scenario }
    \centering
        \caption{Congested scenario }
\label{tab:congested}
    \begin{tabular}{@{}llll@{}}
        \toprule
        {} & DeePC & MPC & No Control \\
        \midrule
        Travel Time (min) & $\boldsymbol{18.41}$ & 20.55 & 20.63\\
        Waiting Time (min) & $\boldsymbol{15.56}$ & 18.35 & 18.43\\
        CO Emissions (kg) & $\boldsymbol{0.168}$ & 0.192 & 0.192\\
        CO$_2$ Emissions (kg) & $\boldsymbol{2.892}$ & 3.241 & 3.254 \\
        HC Emissions ($10^{-4}$kg) & $\boldsymbol{8.35}$ & 9.50 & 9.54 \\
        PMx Emissions ($10^{-4}$kg) & $\boldsymbol{0.70}$ & 0.79 & 0.79 \\
        NOx Emissions ($10^{-4}$kg) & $\boldsymbol{13.13}$ & 14.74 & 14.80 \\
        \bottomrule
    \end{tabular}
\end{table}

\begin{figure}[t]
\centering
     \begin{subfigure}[t]{\columnwidth}
     % \hspace{-0.9cm}
    \includegraphics[width=\textwidth]{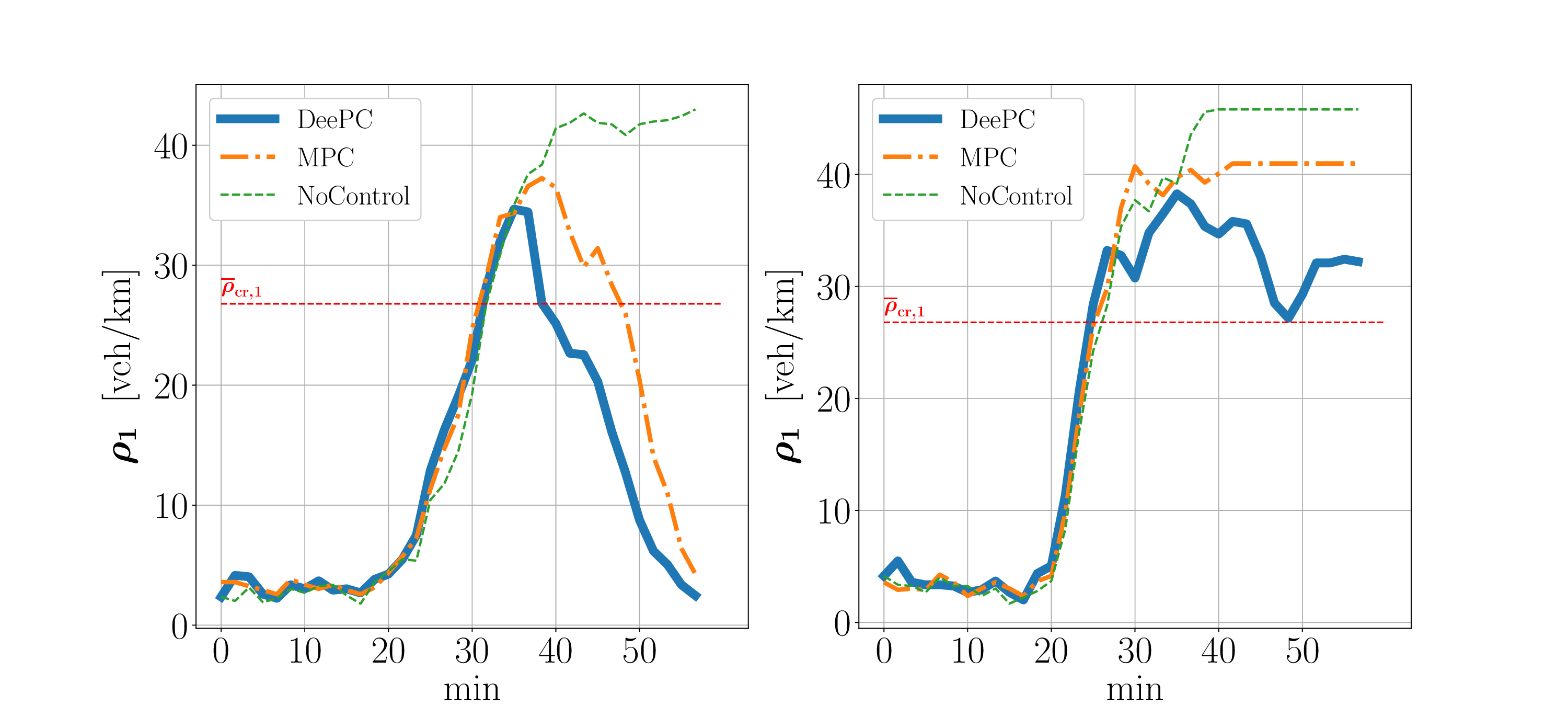}
    \caption{Density region 1}
    \end{subfigure}%
   \\
     \begin{subfigure}[t]{\columnwidth}
     % \hspace{-0.9cm}
    \includegraphics[width=\textwidth]{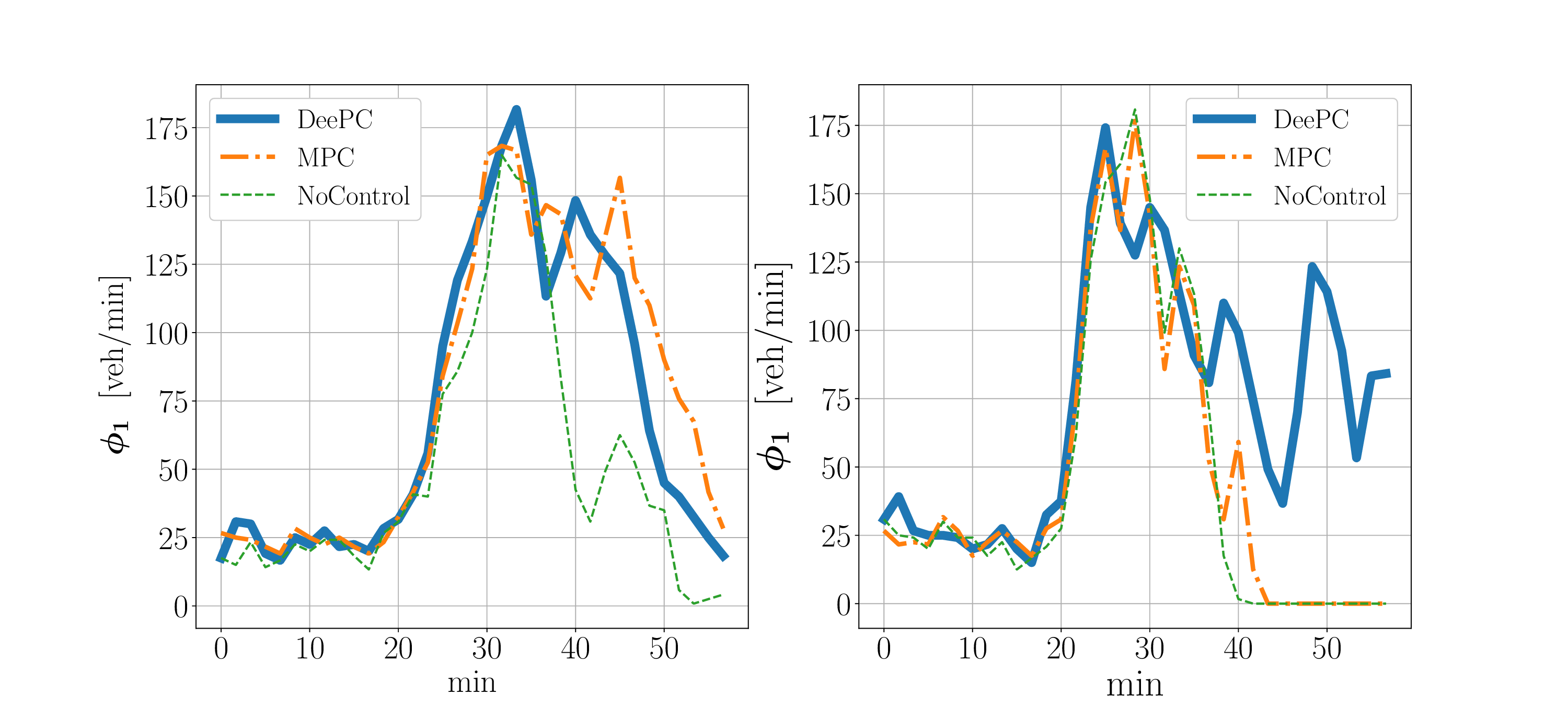}
    \caption{Flow in region 1}
    \end{subfigure}%
   \\
     \begin{subfigure}[t]{\columnwidth}
     % \hspace{-0.9cm}
    \includegraphics[width=\textwidth]{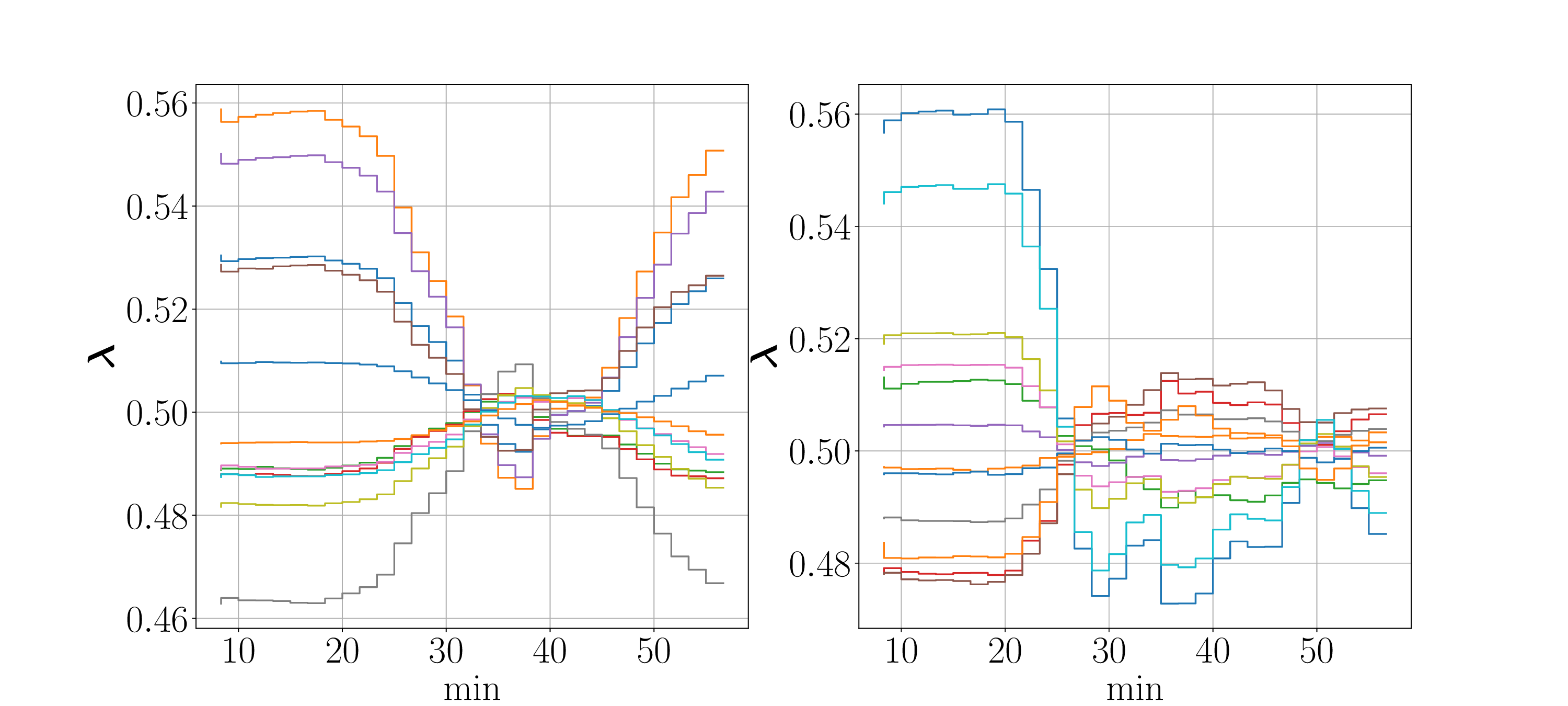}
    \caption{Control input of the traffic lights}
    \end{subfigure}%
   \\
    %  \begin{subfigure}[t]{\columnwidth}
    %  % \hspace{-0.5cm}
    % \includegraphics[width=\textwidth]{./imgs3/error_pyramid.pdf}
    % \caption{Error region 1 uncongested (left) congested (right) }
    % \end{subfigure}%
  % \centering
  \caption{Density (a), flow (b), optimal inputs of DeePC (c) related to the inner region are here shown divided in uncongested (left column) and congested (right column). 
  % Fig. (a)-(b) congested shows that DeePC avoids a gridlock state and maintains flows higher than 0. In Fig. (c) one can observe that in congested states green times are penalized by DeePC.
  } 
  \label{fig:den_flo}
\end{figure}

Tables \ref{tab:uncongested} and \ref{tab:congested} show that \acrshort{DeePC}   outperforms  not having a control and the \acrshort{MPC} baseline both in terms of average travel time  reduction  and CO$_2$ emissions.
\acrshort{DeePC} also decreases the average travel time in the uncongested case by 37.4\% over the no control baseline, compared with the 29.6\% obtained by \acrshort{MPC}.
Fig.\ref{fig:den_flo} (a) highlights that \acrshort{DeePC} maintains the density close to the critical one $\overline{\rho}_{\tup{cr}}$, which in turn maximizes the flow Fig.\ref{fig:den_flo} (a) leading to the lower travel time. 
By taking the average time saved and CO$_2$ emissions and multiplying them for the number of vehicles, we get the total travel time and CO$_2$ that have been saved, respectively 151 hours and 1411 kg for \acrshort{DeePC}. 
In the congested case \acrshort{DeePC} the gain in travel time is 10,7\%, while \acrshort{MPC} does not perform significantly better than the no control baseline. This is because \acrshort{DeePC} prevents the gridlock state, as shown in Fig. \ref{fig:den_flo} (b), and achieves substantially higher flows.
The total travel time and CO$_2$ emissions saved in this scenario are 318 hours and 3113 kg.

\subsection{Case study 2: preliminary analysis of the Z\"urich network}

\begin{figure}[t]
\centering
\includegraphics[width=\columnwidth]{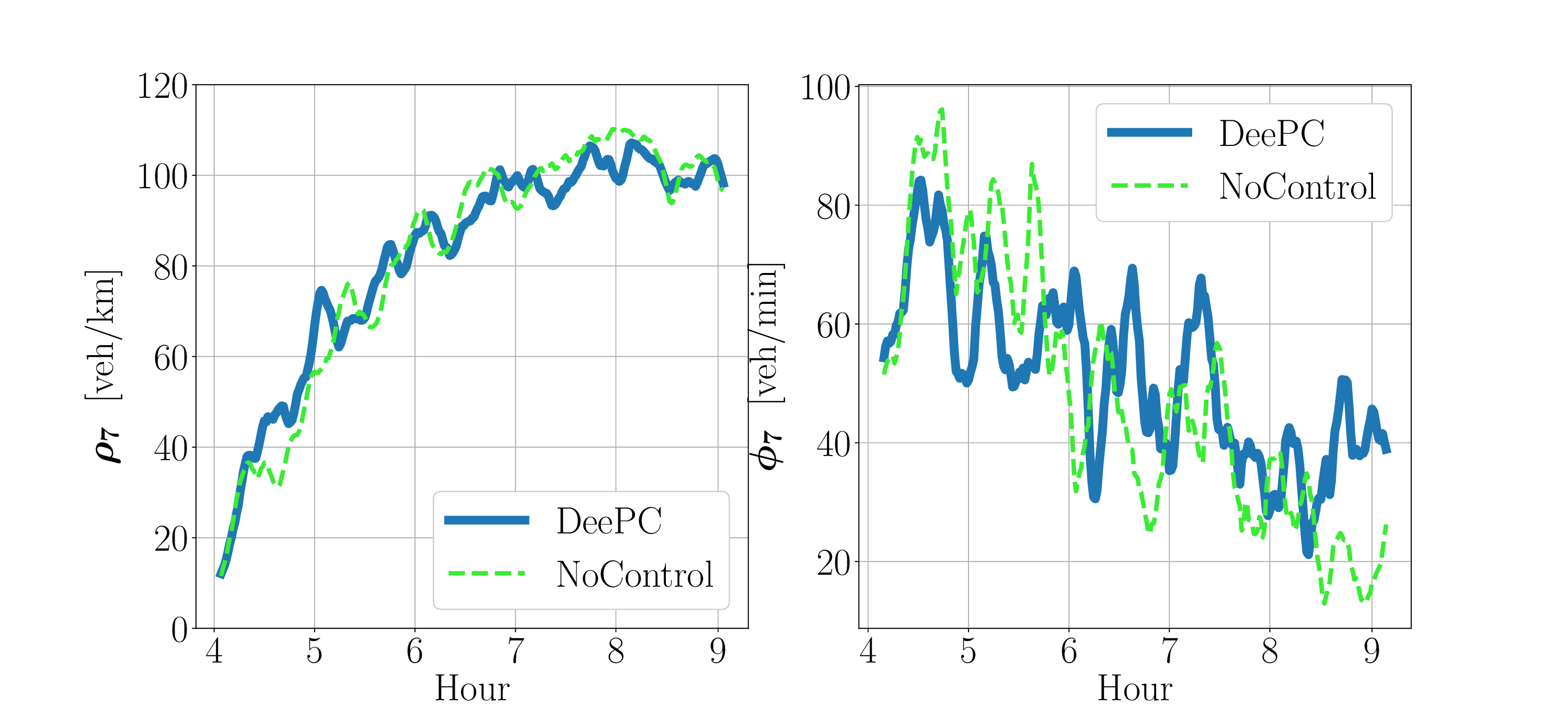}
\centering
\caption{Density (a) and the flow (b) of the center of Z\"urich under a no control policy and the DeePC control.}
\label{fig:Zuri}
\end{figure}

% Building on the results obtained in the lattice network,
We present a preliminary study on  \acrshort{DeePC} over a larger simulation that is the entire traffic network of the city of Z\"urich, Switzerland. 
% Our experiments have been conducted using a high-fidelity mesoscopic SUMO simulation of the entire traffic network of the city of Z\"urich, Switzerland. 
Leveraging a detailed map of the city, counting 14566 roads and 6917 intersections, the simulation offers a good approximation of the urban traffic dynamics by capturing many of the non-linear behaviours observed in reality. Fig.~\ref{fig:Zuri} illustrates the effectiveness of DeePC in controlling the system, resulting in increased flows and a reduced average density in the city center compared to the no-control baseline. This research direction will be explored in greater depth in an upcoming publication.

% \subsection{Numerical example}

% \red{{\bf ``Figures''+ case studies to add}
% \begin{itemize}
%     \item Lattice structure used and divided into regions, highlights the traffic light position
%     \item 2 case studies a congested and uncongested one w. different demands
%     \item Density and flows of the regions w and w/out control
%     \item Table results congested vs uncongested
% \end{itemize}}

\section{Conclusion} \label{sec:conclusion} 
As urban areas continue to expand, exacerbating traffic congestion and emissions, data-driven traffic management strategies offer effective solutions to reduce congestion and emissions. This paper has leveraged the data-driven control algorithm \gls{DeePC} to address the urban traffic control problem.  Our preliminary results, based on a 
case study using the  high-fidelity  SUMO simulation software,  suggests   the great promise of the data-driven approach exemplified by DeePC in outperforming existing methods, particularly in key metrics such as travel time and CO$_2$ emissions. Future research should explore the engineering relevance of this approach with a  real-world   case study.
% and gauge the limitations of this approach by performing a thorough sensitivity analysis of the parameters and data used. 

\bibliographystyle{IEEEtran}
 \bibliography{refs.bib}

\end{document}